\documentclass[%
aps,
rmp,preprint,amsmath,amssymb,graphicx,longbibliography
]{revtex4-1}
 \usepackage{mathptmx}      
\usepackage{latexsym}
\usepackage{amsmath}
\usepackage{epstopdf}
\usepackage{bm}
\usepackage{diagbox}
\usepackage{graphicx}
\usepackage{bm}

\begin{document}

\title{Change in the direction of polarization vector and redshift of an incoming  light ray as observed from a rotating frame.
}

\author{T. Ghosh}
\affiliation{Department of Physics, Assam University, Silchar-788011, Assam, India}
\author{A. K. Sen} 
\affiliation{Department of Physics, Assam University, Silchar-788011, Assam, India}


\begin{abstract}
In the present work, change in the direction of the polarization vector of an incoming  light ray is extensively calculated for a rotating observer. The change in the direction of the polarization vector calculated here is only due to the effect of the non-inertial rotating frame, considering that the light source is at a distance and it is emitting a plane polarized light. The metric tensors for a rotating observer have been collected from existing literature. Accordingly electric displacement and magnetic induction values as applicable for the rotating observer have been calculated. These values are used to calculate the change in the orientation of the electric vector of an incoming plane polarized light ray. Earth has been taken as an example of rotating frame, and the calculated amount of change in the direction of the polarization vector has been found to be dependent on the azimuthal as well as  polar co-ordinate of the rotating frame. Present work also discusses the redshift as observed by a rotating observer and the value of the redshift has been calculated for an observer sitting on rotating earth. 
\end{abstract}

\pacs{Valid PACS appear here}
\keywords{Suggested keywords}
\maketitle

\section{Introduction} \label{sec:intro}
 Light is an electromagnetic wave, and it moves along the curvature of the space-time structure. From this very concept, it is a well-known fact that light shows a unique property, known as the gravitational deflection. Following this idea, it is obvious that the light ray will be deflected, when it passes through a gravitational field, commonly known as gravitational deflection. The gravitational deflection of the light ray gives undeniable proof of the {\it General Theory of Relativity}. Since the establishment of the {\it General Theory of Relativity}, many authors \cite{eddington1919total,dyson1921relativity,balazs1958effect,von1960determination,vilenkin1984cosmic,sereno2004gravitational,sen2010more,roy2015trajectory} have carried out investigations for the proper understanding of the gravitational deflection and the associated polarization phenomena. As gravitational deflection is a well-established phenomenon,  it gives an opportunity to study the polarization state of the light ray which passes through such space-time. Some authors have found in the past \cite{balazs1958effect,1980ApJ...238.1111S} that when a light ray passes through gravitational field having axial symmetry, then its polarization vector is rotated by the gravitational field, quite analogous to the phenomenon which is called Faraday Rotation by the magnetic field. More recently \cite{ghosh2016effect} investigated the change in the direction of polarization vector in the case of both Schwarzschild and Kerr field. They showed that although the Schwarzschild field does not affect the state of polarization of light, but Kerr field produces a change in the direction of the polarization vector of the electromagnetic wave or light ray. Though there are reported works on the change in the direction of the polarization vector by the space-time geometry introduced by gravitational mass, there is not much investigation conducted in case of accelerated systems. Thus a natural question lies here, whether the space-time transformation generated by an accelerated observer has any effect on the polarization state of the light ray or not? Here in the present work, the space-time transformation generated by a rotating frame is the subject of investigation. \\
 \\
In a discuss ion on the accelerated system, \cite{mashhoon1990hypothesis} raised the question about the law which specifies the measurement by an accelerated observer and focused on the hypothesis of locality which is described as {\it ``The presumed equivalence of an accelerated observer with a momentarily co-moving inertial observer-underlies the standard relativistic formalism by relating the measurements of an accelerated observer to those of an inertial observer''}. \cite{mashhoon1990hypothesis} discussed briefly, the significance and the limitations of the hypothesis of locality in his work. To understand the motion under the influence of accelerated frame one must consider the hyperbolic motion in space-time, which had been elaborately discussed by \cite{rindler1960hyperbolic}. He was the first person who studied relativistic motion under an accelerated system and later Born coined it as “hyperbolic motion”. \cite{rindler1960hyperbolic} had obtained the differential equation of motion for a particle in uniform accelerated frame generalising the geometric characteristics of a rectangular hyperbola in Minkowskian space-time.  It is termed as “Hyperbolic motion” from the fact that, as seen by the inertial frame observer if we plot the distance against the time on a Minkowski diagram, it describes a hyperbola.\\  
In case of inertial frames,  we can transform any four-vector by performing a multiplication operating  by a 4x4 Lorentz transformation matrix. The electric and magnetic  field vectors are components of a field tensor $(F^{ik})$ of  rank two. Thus by operating with Lorentz Matrix twice we can transform electric and magnetic   vectors between any two inertial frames\cite{landau1971classical} [page no.
66].\\
However, when we have linear or rotational acceleration, we can not identify any Lorentz like transformation Matrix to perform similar co-ordinate transformation.
Let us consider two reference frames $K^{\prime \prime} (x^{\prime \prime0}, x^{\prime \prime1}, x^{\prime \prime2}, x^{\prime \prime3})$ and $K (x^{0}, x^{1}, x^{2}, x^{3})$ representing  accelerated (non-inertial) and inertial frame respectively. Under such a situation the transformation relation can be given  by Equations (15a) and (15b) of Nelson (1987).
Since this type of transformation relation can not be replaced by a Lorentz like transformation, so it is more convenient to discuss the transformation of electric and magnetic field vectors, using  suitable metric tensors '$g_{ik}$'\cite{nelson1987generalized,torres2006uniformly}\\
To find out the coordinate transformation relation for the linearly accelerated frame let us consider $a$, as the proper linear acceleration of the observer, and it is accelerated parallel to the $+x^{1}$ axis. In {\it  General Relativity}, proper acceleration is an acceleration measurable by an accelerometer rigidly tied to a frame and does not occur by gravitation. We consider that an observer is accelerated and its coordinate system is ${\bf \tilde{K}(\tilde x^{0}, \tilde x^{1}, \tilde x^{2}, \tilde x^{3})}$. Then ${\bf K( x^{0}, x^{1}, x^{2}, x^{3})}$ defines the coordinate frame where the inertial source is situated.\\
Now for the constant acceleration, ($a$) the Rindler co-ordinate for the accelerated observer in instantaneous inertial frame $\tilde{K}$ (as defined above) in terms of the co-ordinate of the  inertial frame $K$ can be written as \cite{torres2006uniformly,rindler2012essential,carroll2004spacetime,stephani2004relativity,misner2017gravitation};
\begin{eqnarray}    
\begin{array}{lcl}
x^{0}=(\frac{c^{2}}{a}+\tilde{x}^{1})\sinh(\frac{a\tilde{x}^{0}}{c^{2}})\\
x^{1}=(\frac{c^{2}}{a}+\tilde{x}^{1})\cosh(\frac{\tilde{x}^{0}}{c^{2}}).
\end{array}
\label{Acceleration_7}
\end{eqnarray}
This is the most general form of Rindler coordinates for a uniform accelerated observer. \cite{nelson1987generalized} also derived the transformation equation for a non-rotational uniform linear accelerated frame. He also derived the equation for the motion by the time-dependent, non-gravitational acceleration. Later \cite{nelson1987generalized} extensively worked on the transformation relation between the rest frame and the accelerated frame considering both linear acceleration and rotational acceleration. Later his work had shown that the relation satisfies the familiar Rindler metric also. A similar transformation relation can be obtained from work by \cite{alsing2004simplified}. In their work \cite{alsing2004simplified} briefly discussed Hawking-Unruh radiation temperature for an accelerated frame. Also from their work, the transformation relation for the uniform accelerated frame can be derived as in Eqn.\ref{Acceleration_7}. Recently the authors \cite{scarr2016solutions} had obtained the velocity as well as coordinate transformation relation for a uniformly accelerated frame. In one of their previous work, they \cite{friedman2013spacetime} had discussed the four-dimensional covariant relativistic equation and discussed the concept of a maximal acceleration. \cite{torres2006uniformly} studied the uniformly accelerated motion, and estimated the red-shift of an electromagnetic wave.\\
\cite{mashhoon1990hypothesis} showed that an observer  rotating in a frame  with radius $r$ has transnational acceleration $r\Omega^{2}\gamma^{2}$, where $\gamma$ is the Lorentz factor, and $\Omega$ is the rate of rotation of frame per unit time (in other words angular velocity). Thus in continuation of the work by him, the rotating observer can be considered as an accelerated system, and the metric for the system has been given by \cite{mashhoon1990hypothesis};
\begin{eqnarray}    
\begin{array}{lcl}
g_{00}=\gamma^{2}[1-\frac{\Omega^{2}}{c^{2}}(r+x^{1})^{2}-\frac{\Omega^{2}\gamma^{2}}{c^{2}}(x^{2})^{2}]\\
g_{0\alpha }=-[\frac{\Omega\gamma^{2} }{c}\times \vec{X}]\\
g_{\alpha \beta}=-\delta_{\alpha \beta}.
\end{array}
\label{Acceleration_8}
\end{eqnarray}
As per \cite{mashhoon1990hypothesis}, the rotating observer can be characterised by proper acceleration length $\frac{c}{\gamma \Omega}$, where 
the above $X=(x^{1}, x^{2}, x^{3})$ vector.\\
\cite{adams1925relativity} first confirmed gravitational redshift from the measurement of the apparent radial velocity of $Sirius$ $B$, since then many 
observations were made to measure the gravitational red-shift \cite{PhysRevLett.3.9,PhysRevLett.28.853}. More recently, \cite{PhysRevLett.70.2213} measured 
the gravitational redshift of the sun in the year 1993. Recently \cite{dubey2015analysis} gave the analytic result for the gravitational redshift observed from a 
rotating body and as observed by an asymptotic observer. The authors also calculated the numerical value of redshift of some rotating heavenly bodies. 
In section \ref{subsection:redshift}, of the present work  the redshift, $Z$ as observed by an rotating observer for light coming from an inertial frame  has been 
calculated analytically along with their numerical values and for that matter the observer was considered on the surface  of the earth.


\section{Discussion on rotating observer } \label{sec:rotatingmetric}
  Now let us consider the observer is sitting on a rotating frame, which has the rate of rotation $\Omega$. According to \cite{mashhoon1990hypothesis} 
  the distinction between accelerated observer in Minkowskian space-time and co-moving inertial observer is the presence of acceleration scales associated with non-inertial 
  observer. In this paper, the rotation is considered around with the $x^{3}$ or z-axis. The light source is situated in a space where frame-dragging and other 
  gravitational effect of rotating frame (body) is negligible. Further the light ray is coming along $x^{1}$ or x-axis of the Cartesian system. 
  From Eqn.\ref{Acceleration_8} the metric elements for the rotating frames can be derived, and these have been given by \cite{mashhoon1990hypothesis};
\begin{eqnarray}    
\begin{array}{lcl}
g_{00}=\gamma^{2}[1-\frac{\Omega^{2}}{c^{2}}(r+x^{1})^{2}-\frac{\Omega^{2}\gamma^{2}}{c^{2}}(x^{2})^{2}]\\
g_{01}=\frac{\gamma^{2}\Omega}{c} (x^{2})\\
g_{02}=-\frac{\gamma^{2}\Omega}{c}  (x^{1})\\
g_{03}=0\\
g_{\alpha \beta}=-\delta_{\alpha \beta}.
\end{array}
\label{Acceleration_9}
\end{eqnarray}
As the components of $\Omega$ lies only along $x^{3}$ or z-axis exists, the other components of $\Omega$ don't exist (i.e. $\Omega_{x}=\Omega_{y}=0$ and $\Omega_{z}=\Omega$).
Again, if the rate of rotation $\Omega $ becomes zero then from the Eqn.\ref{Acceleration_9}, the metric elements are given as $g_{00}=1, g_{0\alpha}=0, g_{\alpha \beta}
=-\delta_{\alpha \beta}$, which are the metric for flat space-time.\\

\subsection{Change in polarization vector for a general case} \label{subsec:polarization}
 In order to calculate the amount by which the polarization vector of the light ray will be changed (rotated), we consider the light ray from the source to be plane polarized 
 (as was done in our earlier work \cite{ghosh2016effect}). To determine the position angle of the polarization vector of light ray received by the observer from the distant source, 
 one has to consider the initial components of the polarization vector of the light ray $E_{y}$ and $E_{z}$ along the direction of $x^{2}$ and $x^{3}$-axis respectively, where one 
 should recall the fact that light ray has been travelling along $x^{1}$-axis. Let us assume that the magnitude ratio between $E_{y}$ and $E_{z}$ is $\xi$, in other
 words $E_{y}=\xi E_{z}$ \cite{ghosh2016effect}. So the initial position angle of the polarization vector $\chi_{source}$ is given by;
\begin{equation}\label{Acceleration_10}
  \chi_{source}=\arctan(\frac{E_{y}}{E_{z}})=\arctan \xi
\end{equation}
Now the position angle of the polarization vector as will be measured by the observer in the rotating frame will be \cite{ghosh2016effect}:
\begin{equation}\label{Acceleration_11}
  \chi_{observer}=\arctan(\frac{D_{y}}{D_{z}})
\end{equation}
Where $D_{y}$ and $D_{z}$ are the electric displacement component along the $x^{2}$ and $x^{3}$-axis respectively. To find out the value of $\chi_{observer}$ one must find out the 
relationship between electric displacement vector $\vec{D}$ and electric vector $\vec{E}$. The relation between electric displacement vector and the electric vector has been given 
by \cite{landau1971classical}:
\begin{equation}\label{Acceleration_12}
  \vec{D}=\frac{\vec{E}}{\sqrt{g_{00}}} + \vec{H} \times \vec{g}
\end{equation}
Where $\vec{H}$ is the magnetic induction of the light ray received. Now from the Eqn.\ref{Acceleration_12} the components $D_{y}$ and $D_{z}$ are deduced 
as\cite{landau1971classical}, \cite{ghosh2016effect}:
\begin{eqnarray}    
\begin{array}{lcl}
D_{y}=\frac{E_{y}}{\sqrt{g_{00}}}+H_{z}g_{01}\\
D_{z}=\frac{E_{z}}{\sqrt{g_{00}}}-H_{y}g_{01}
\end{array}
\label{Acceleration_13}
\end{eqnarray}
The different components of the magnetic field of the incoming light ray can be written in terms of the electric field using the 
relation $H_{z}=|\vec{H_{z}}|=|\hat{n} \times \vec{E_{y}}|$ \cite{landau1971classical} [page no. 112], where $\hat{n}$ is the unit propagation vector
parallel to the propagation of light ray (eg. parallel to x-axis). Now taking the note that $E_{y}=\xi E_{z}$ and considering only the magnitude of $\vec{H}$,
we can write from Eqn.\ref{Acceleration_13}:
\begin{eqnarray}  \label{Acceleration_14}  
\begin{array}{lcl}
D_{y}=E_{z}\frac{\xi}{\sqrt{g_{00}}}+\xi g_{01}\\
D_{z}=E_{z}\frac{1}{\sqrt{g_{00}}}- g_{01}\\
\end{array}
\end{eqnarray}\\
So from Eqn. \ref{Acceleration_11} with the help from Eqn. \ref{Acceleration_14} it can be written as \cite{ghosh2016effect}:
\begin{equation}    
\chi_{observer}=\arctan(\xi \frac{1 + \sqrt{g_{00}}g_{01}}{1- \sqrt{g_{00}}g_{01}})
\label{Acceleration_15}
\end{equation}
So the total change in the direction of polarization vector of received light ray (say $\chi$) can be calculated from Eqn.\ref{Acceleration_15} and Eqn.\ref{Acceleration_10}, as:
\begin{eqnarray}\label{Acceleration_16_1}
 \nonumber 
  \chi &=& \chi_{observer}-\chi_{source} \\
  \nonumber
   &=& \arctan (\xi \frac{1 + (\sqrt{g_{00}})g_{01}}{1- (\sqrt{g_{00}})g_{01}})-\arctan \xi \\
   &=& \arctan \left(\frac{2\xi (\sqrt[]{g_{00}})g_{01}}{1+\xi^{2}+\xi^{2}(\sqrt[]{g_{00}})g_{01}-(\sqrt[]{g_{00}})g_{01}} \right)
\end{eqnarray}
It is clear from Eqn.\ref{Acceleration_16_1} that the total change in the direction of polarization vector of light ray  depends on the initial position angle of the polarization vector of the light ray emitted from a source. Thus Eqn. \ref{Acceleration_16_1} gives the change in the direction of polarization vector of light ray received by a rotating observer (i.e non-inertial frame).\\
Now, for the simplicity of the problem let us consider, $E_{y}=E_{z}$, (which indicates the polarization vector is held at an angle $45^\circ$  with respect to the rotation axis). Then the change in the direction of polarization vector can be given as:
\begin{equation}
\label{Acceleration_16}
\chi= \arctan \left( \sqrt[]{g_{00}}g_{01} \right)
\end{equation}
\subsection{Change in the direction of polarization vector as seen from rotating frame} \label{subsec:rotationofpolarization}
To determine the total change in the direction of polarization vector by a rotating observer, one should recall the metric given in Eqn.\ref{Acceleration_9}. 
From subsection \ref{subsec:polarization}, Eqn. \ref{Acceleration_16}, it is clear that only the metric components $g_{00}$, and $g_{01}$ given in Eqn.\ref{Acceleration_9}, are 
necessary to determine the total change in the direction of polarization vector as observed by an observer, who is placed on an  accelerated non-inertial frame. 
Now in Eqn. \ref{Acceleration_9} the Lorentz factor $ \gamma $ has been introduced by\cite{mashhoon1990hypothesis};
\begin{equation}\label{Acceleration_17}
  \gamma=\frac{1}{\sqrt{1-(\frac{\Omega r}{c})^{2}}}
\end{equation}
Let us assume that the observer is situated at the azimuthal position $ \phi $ and polar position $\theta $ on the surface of a rotating frame of radius $r$, so the Cartesian 
y co-ordinate of the observer is given by $(x^{2})=r \sin (\theta) \sin (\phi) $ and x co-ordinate is given by $(x^{1})=r \sin (\theta) \cos (\phi)$. Now with the help of 
Eqn.\ref{Acceleration_9}, Eqn. \ref{Acceleration_16} and Eqn. \ref{Acceleration_17} the total change in the direction of polarization vector for a rotating observer can be written as;
\begin{eqnarray}\label{Acceleration_18}
\nonumber
  \chi &=& \arctan (\sqrt{g_{00}}g_{01})\\
\nonumber
        &=& \arctan \left[ \sqrt{\left( \gamma^{2}[1-\frac{\Omega^{2}}{c^{2}}(r+x^{1})^{2}-\frac{\Omega^{2}\gamma^{2}}{c^{2}}(x^{2})^{2}] \right)} \left( \frac{\gamma^{2}\Omega}{c} (x^{2}) \right) \right]\\
\nonumber
        &=& \arctan \left [
        \sqrt{\left(
            \frac{1}{1-(\frac{\Omega r}{c})^{2}}[1-\frac{\Omega^{2}}{c^{2}}(r+r \sin \theta \cos \phi)^{2}-\frac{\Omega^{2}\gamma^{2}}{c^{2}}(r \sin \theta \sin \phi)^{2}]
        \right)}\left(
                    \frac{1}{1-(\frac{\Omega r}{c})^{2}}\frac{\Omega r\sin \theta \sin \phi}{c}
                \right)
        \right] \\
  &=&\arctan \left[ \sqrt{\frac{c^{2}}{c^{2}-(\Omega r)^{2}}\left\{[1-\frac{\Omega^{2}}{c^{2}}(r+r\sin \theta \cos \phi)^{2}-\frac{(c\Omega r)^{2}}{c^{2}-(\Omega r)^{2}}\sin^{2} \theta \sin^{2}\phi\right\}}\left(\frac{c\Omega r\sin \theta \sin \phi}{c^{2}-(\Omega r)^{2}}\right) \right]
\end{eqnarray}
Eqn. \ref{Acceleration_18} gives the total change in the direction of polarization vector of light ray as observed  by a rotating observer situated on the surface of a frame of radius $r$, 
with azimuthal position $\phi $ and polar position $\theta $ having rate of rotation $\Omega $. These co-ordinate $\phi $ and $\theta $ can be identified as longitude and 
( $90^{\circ}$ - latitude).
\subsubsection{Case 1: For $\theta = 0 $, polar position of observer} \label{subsubsection:casepi}
Now, consider a case when the polar position of the observer is given by $\theta =0 $ (observer is sitting on the north pole of rotating frame), and with the help of
Eqn. \ref{Acceleration_18} it is clear that there is no change in the direction of polarization vector of light ray received by a rotating observer. 
Putting $\theta = 0 $ the $g_{01}$ component in Eqn. \ref{Acceleration_18} becomes zero which in turn makes $\chi =0 $ (i.e. there is no change in the direction of polarization vector). 
In case of earth the pole is represented by $90^{o}$ latitude, where $\theta =0 $ and the relation between earth latitude (say $\Theta $) is $\Theta = (90^{o}-\theta ) $. So from above 
discussion, it is clear that there would be no change in the direction of polarization vector if the light ray is observed  from any of the two poles of the earth.
\subsubsection{Case 2: For $\theta = \frac{\pi}{2} $ equatorial position of observer} \label{subsubsection:case0}
\begin{figure}[!htb]
\centerline{\includegraphics[width=9cm]{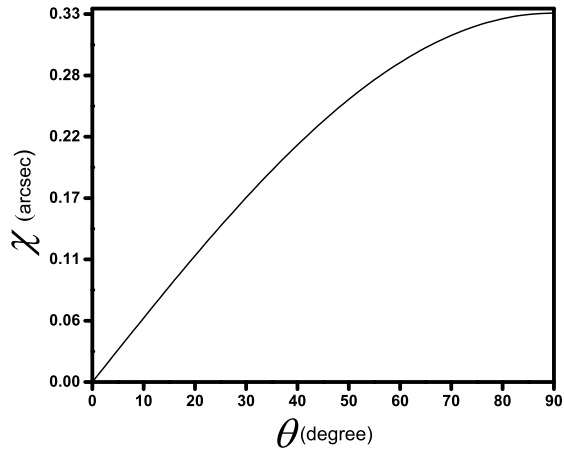}}
\caption{Variation of total change in the direction of polarization vector, $\chi$ (in arcsec) versus the polar position, $\theta$ (in degree), where $\phi$ is kept constant at $\frac{\pi}{2}$. 
\label{fig:earththeta}}
\end{figure}

In this case, let us consider the polar position of the rotating observer is given by $\theta= \frac{\pi}{2} $, then with the help of Eqn. \ref{Acceleration_18} the total rotation of 
the polarization vector can be written as;
\begin{equation}\label{Acceleration_19}
   \chi_{\theta=\frac{\pi}{2}}=\arctan \left[
        \sqrt{\left \{
            \frac{c^{2}}{c^{2}-(\Omega r)^{2}}\left(
                    1-\frac{r^{2}\Omega^{2}}{c^{2}}(1+\cos \phi )^{2}-\frac{(c\Omega r)^{2}}{c^{2}-(\Omega r)^{2}}\sin^{2} \phi
                    \right)
           \right \}}\left(
                     \frac{c\Omega r\sin \phi }{c^{2}-(\Omega r)^{2}}
                     \right)
                                     \right]
\end{equation}
 Now from Eqn.\ref{Acceleration_19} one can observe that, $\chi_{\theta =\frac{\pi}{2}} $ (observer on the equator) depends on the azimuthal position ($\phi $, equivalent to earth's 
 longitude) of the observer. In case $\phi = 0$ there will be no change in the position of polarization vector (as $\sin 0=0$) and the rotation is maximum when 
 $\phi = \frac{\pi}{2}$ or $-\frac{\pi}{2}$. So the maximum change in the direction of polarization vector could be observed when the rotating observer is at azimuthal 
 position $\phi $ = $\frac{\pi}{2}$ or $-\frac{\pi}{2}$  and polar position $\theta = \frac{\pi}{2}$. Now for the position $\phi =\frac{\pi}{2}$ the value of $\chi $ is +ve and 
 for $\phi =-\frac{\pi}{2}$ $\chi $ is -ve. These two positions represent the prograde and retrograde cases respectively. The numerical value of $\chi $ in both cases are same,
 but they are opposite in sign (i.e. $\chi_{\theta=\phi=\frac{\pi}{2}}=-\chi_{\theta=\frac{\pi}{2},\phi=-\frac{\pi}{2}}$). So the maximum change in the direction of polarization 
 vector is given by;
\begin{equation}\label{Acceleration_20}
  \chi_{\theta=\phi=\frac{\pi}{2}}=\arctan  \left[
    \sqrt{\left \{
        \frac{c^{2}}{c^{2}-(\Omega r)^{2}}\left(
            1-\frac{\Omega^{2}}{c^{2}}(r )^{2}-\frac{(c\Omega r)^{2}}{c^{2}-(\Omega r)^{2}}
            \right)
        \right \}}\left(
            \frac{c\Omega r}{c^{2}-(\Omega r)^{2}}
                \right)
                                            \right]
\end{equation}
 The rate of rotation of earth, $\Omega_{earth}$ is $7.27 \times 10^{-5}$ rad/sec and the average radius of earth is $r_{earth}$ is $63.78 \times 10^{5}$ m. On 
 the equatorial plane the latitude of the earth is $0$ and, for the azimuthal position, $\phi = \frac{\pi}{2}$ the longitude is given by $90^{\circ}$. 
 From Eqn. \ref{Acceleration_20} the total change in the direction of polarization vector of light ray, received by an observer on the equatorial position 
 can be calculated as  $\chi_{earth,\theta=\phi=\frac{\pi}{2}}= 0.33$ arcsec. Similarly the total change in the direction of polarization vector at longitude $90^{\circ}$ west 
 is, $\chi_{earth,\theta=\frac{\pi}{2},\phi=-\frac{\pi}{2}}=-0.33$ arcsec. Now, let us consider an extreme  case where $\Omega r \simeq c$, the velocity of the light ray.
 In that case, Lorentz factor $\gamma $ becomes undefined, so the metric elements $g_{00}$, $g_{01}$ and $g_{02}$ also become undefined, as can be seen from Eqn.\ref{Acceleration_9}. 
 Thus the value of $\chi$ cannot be calculated.\\
\begin{figure}[!htb]
\centerline{\includegraphics[width=9cm]{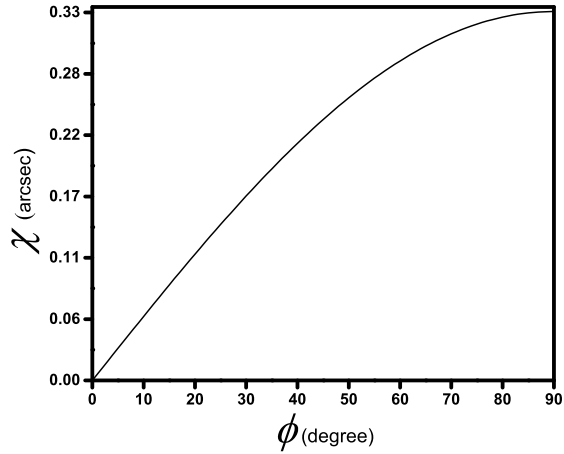}}
\caption{Variation of total change in the direction of polarization vector, $\chi$ (in arcsec) at the equator ($\theta=\frac{\pi}{2}$) vs the azimuthal position, $\phi$ (in degree)
\label{fig:earthphi}}
\end{figure}
The total change in the direction of polarization vector, $\chi$ as observed  by a rotating observer at equator of earth  versus the azimuthal position ($\phi$) of the observer is 
shown in Fig.\ref{fig:earthphi}. Again, in the case when $\phi=\frac{\pi}{2}$ and the total change in the direction of polarization vector, $\chi$ as a function of  the polar position
of the observer has been shown in the Fig. \ref{fig:earththeta}.
\\
From both the figures Fig.\ref{fig:earthphi} and Fig.\ref{fig:earththeta} one can observe that $\chi$ varies in a similar fashion for both the azimuthal and the polar position. It 
attains a maximum value when $\theta=\phi=\frac{\pi}{2}$, as mentioned before.
 This value of rotation which is about 0.33 arc sec, can be measured with a good polarimeter. In principle, by sitting on earth and by making successive measurements at dawn,
 mid-noon and dusk, we can verify our calculations. The variation of total change in the direction of polarization vector,  against  azimuthal position $(\phi)$ and 
 polar position $(\theta)$ is shown, taking earth as an example of rotating frame in Table 1
\begin{center}
\begin{table}
\caption{change in the direction of polarization vector $\chi$, in arcsec}
\begin{tabular} 
{|l||*{7}{c|}}\hline
\backslashbox[1.0 cm]{\!$\! \theta$}{$\phi$}
&\makebox[1.1 cm]{0}&\makebox[1.1 cm]{15}&\makebox[1.1 cm]{30}
&\makebox[1.1 cm]{45}&\makebox[1.1 cm]{60}&\makebox[1.1 cm]{75}&\makebox[1.1 cm]{90}\\\hline\hline
0 &0.0000&0.0000&0.0000&0.0000&0.0000&0.0000&0.0000\\\hline
15 &0.0000&0.0222&0.0428&0.0606&0.0742&0.0827&0.0856\\\hline
30 &0.0000&0.0428&0.0827&0.1170&0.1433&0.1598&0.1655\\\hline
45 &0.0000&0.0606&0.1170&0.1655&0.2027&0.2260&0.2340\\\hline
60 &0.0000&0.0742&0.1433&0.0202&0.2482&0.2768&0.2866\\\hline
75 &0.0000&0.0827&0.1598&0.2260&0.2768&0.3088&0.3197\\\hline
90 &0.0000&0.0856&0.1655&0.2340&0.2866&0.3197&0.3309\\\hline
\end{tabular}
\end{table}
\end{center}


\subsection{A different case: When the rotation axis of the observer is along the direction of propagation of light} \label{subsection_another}
Now, consider a case when the light is coming parallel to the rotation axis of the observer's the rotational frame. In this case, the only component of the 
rotational vector (of the frame) exists along $x^{1}$-axis. To calculate the change in the direction of polarization vector $\chi $, the values of $g_{00}$ and $g_{01}$ have 
to be calculated, from Eqn.\ref{Acceleration_16}.  The value of $g_{01}$ can be calculated from Eqn.\ref{Acceleration_9} which is,
\begin{equation} \label{Another_1}
    g_{01}= \frac{r^{2}\Omega^{2}}{c^{2}}\cdot \left(0\right)=0
\end{equation}
So, from Eqn.\ref{Acceleration_16} it is clear that there will be no change in the direction of polarization vector if the light ray is coming parallel to the rotation axis 
of the frame and we will get, $\chi =0$.

\section{Discussion on observer who is sitting on an uniform rectilinear accelerated frame}\label{section:rectilinearacceleration}
Now, suppose that the observer's frame is moving with a uniform rectilinear acceleration along the positive $x^{1}$ direction and situated at 
$K^{\prime}(x^{0 \prime },x^{1 \prime },x^{2 \prime },x^{3 \prime },)$ frame. Then there will be  no rotation. Light ray is emitted from a inertial frame given by 
frame $K(x^{0},x^{1},x^{2},x^{3})$. In this case the the transformation equation will be given by \cite{rindler1960hyperbolic}, in the present Eqn.\ref{Acceleration_7}. 
From this equation \ref{Acceleration_7} we can write the line element as:
\begin{equation}
\label{linear_1}
ds^{2}=(1+\frac{a x^{1\prime}}{c^{2}})dx^{0\prime}-(dx^{1\prime})^{2}-(dx^{2\prime})^{2}-(dx^{3\prime})^{2}
\end{equation}
From Eqn.\ref{linear_1} it is clear that the metric element $g_{01}$ is zero, hence  with the help of Eqn.\ref{Acceleration_12} one can obtain the displacement 
vector $\vec{D}=\frac{\vec{E}}{\sqrt[]{g_{00}}}$. Now the two components of displacement vector along $x^{2\prime}$ and $x^{3\prime}$ directions are given as:
\begin{eqnarray}    
\begin{array}{lcl}
D^{\prime}_{y}=\frac{E_{y}}{\sqrt{g_{00}}}\\
D^{\prime}_{z}=\frac{E_{z}}{\sqrt{g_{00}}}
\end{array}
\label{linear_2}
\end{eqnarray}
Here $D^{\prime}_{y}$ and $D^{\prime}_{z}$ are the displacement vector received by the observer situated on $K^{\prime}$ frame. So the position angle of the polarization vector 
of the received ray would be :
\begin{equation}
\chi^{\prime}_{observer}=\arctan (\frac{D^{\prime}_{y}}{D^{\prime}_{z}})=\arctan (\frac{E_{y}}{E_{z}})=\chi_{source}
\end{equation}
where $\chi^{\prime}_{observer}$ is the position angle of the polarization vector as observed by the receiver and $\chi_{source}$ is the position angle of the polarization vector 
of emitted light ray. $\chi_{source}$ is given in the earlier as $\arctan(\frac{E_{y}}{E_{z}})$. It is clear from Eqn.(\ref{linear_2}) that there will be no change in the direction 
of polarization vector for this case. 

\section{Redshift caused by the rotating frame}\label{subsection:redshift}
To calculate the redshift when the light is emitted by  a distant inertial source and observed  by a rotational ( non-inertial)  observer, the four-velocity of the observer has to be 
determined, and it is obtained as \cite[page no. 23]{landau1971classical}:
\begin{equation}\label{red_1}
  u^{i}=\frac{dx^{i}}{ds}
\end{equation}
From Eqn. \ref{Acceleration_9} the line element for the rotational observer can be written as:
\begin{equation}\label{red_2}
  ds=\sqrt{g_{00}(dx^{0})^{2}+g_{01}dx^{0}dx^{1}+g_{02}dx^{0}dx^{2}+\delta_{\alpha \beta} g_{\alpha \beta}dx^{\alpha}dx^{\beta}}
\end{equation}
The values of $g_{00}$, $g_{0x}$, or $g_{01}$ and $g_{xx}$ or $g_{11}$ etc. have been calculated earlier Eqn.\ref{Acceleration_9}, for the co-ordinate 
system $(x^{0},x^{1},x^{2},x^{3})=(ct,x,y,z)$. Now for our convenience in the present geometry, we can  rewrite the metric tensors values in terms of $( ct, r, \theta, \phi)$, 
which are as follows :
\begin{eqnarray}    
\begin{array}{lcl}
g_{00}=\left(
            \frac{1}{1-(\frac{\Omega r}{c})^{2}}[1-\frac{\Omega^{2}}{c^{2}}(r+r \sin \theta \cos \phi)^{2}-\frac{\Omega^{2}\gamma^{2}}{c^{2}}(r \sin \theta \sin \phi)^{2}]
        \right)\\
g_{01}=\left(
                    \frac{1}{1-(\frac{\Omega r}{c})^{2}}\frac{\Omega r\sin \theta \sin \phi}{c}
                \right)\\
g_{02}=-\left(
                    \frac{1}{1-(\frac{\Omega r}{c})^{2}}\frac{\Omega r\sin \theta \cos \phi}{c}
                \right)\\
g_{03}=0\\
g_{\alpha \beta}=-\delta_{\alpha \beta}.
\end{array}
\label{Red_1}
\end{eqnarray}
Thus we may write:
\begin{eqnarray} 
\nonumber
ds^{2} &=& \left(
            \frac{1}{1-(\frac{\Omega r}{c})^{2}}[1-\frac{\Omega^{2}}{c^{2}}(r+r \sin \theta \cos \phi)^{2}-\frac{\Omega^{2}\gamma^{2}}{c^{2}}(r \sin \theta \sin \phi)^{2}]
        \right)(cdt)^{2} \\
       & &-\left(
                    \frac{\Omega r\sin \theta \sin \phi}{1-(\frac{\Omega r}{c})^{2}}\frac{1}{c}
                \right)r \sin \theta \sin \phi d\phi dt+\left(
                    \frac{\Omega r\sin \theta \cos \phi}{1-(\frac{\Omega r}{c})^{2}}\frac{1}{c}
                \right)r \sin \theta \cos \phi d\phi dt
\label{Red_1}
\end{eqnarray}
For any calculation of redshift introduced by space-time, we must begin with the invariant.
\begin{equation}
    [k_{i}u^{i}]_{source} =[k_{i}u^{i}]_{observer}
\end{equation}
Now, the components of four velocity can be written as:
\begin{eqnarray}    
\begin{array}{lcl}
  u^{0}=\frac{dx^{0}}{ds}=\frac{cdt}{ds}\\
  u^{1}=\frac{dx^{1}}{ds}\\
  u^{2}=\frac{dx^{2}}{ds}\\
  u^{3}=\frac{dx^{3}}{ds}
\end{array}
\label{red_3}
\end{eqnarray}
Now let $\Gamma$ be any arbitrary quantity which describes the wave field. For a plane monochromatic wave $\Gamma $ can be written as\cite{landau1971classical} [page no. 140]:
\begin{equation}\label{red_4}
  \Gamma=a e^{\textbf{j}(k_{i}x^{i})+\alpha}
\end{equation}
Where {\bf j=$\sqrt{-1}$,} $k_{i}$ is wave four vector. One can write the expression for the field as, $\Gamma=a e^{\textbf{j}\Psi}$. $\Psi$ is the eikonal. 
Over small region of space and time intervals, the eikonal, $\Psi$ can be expanded in series to terms of the first order and it is written as:
\begin{equation}\label{red_5}
  \Psi=\Psi_{0}+r\frac{\partial \Psi}{\partial r}+ t \frac{\partial \Psi}{\partial t}
\end{equation}
From this,  one can have\cite{landau1971classical}
\begin{equation}\label{red_6}
  k_{i}=-\frac{\partial \Psi}{\partial x^{i}}
\end{equation}
Again it can be shown that\cite{landau1971classical},
\begin{equation} \label{red_7}
k^{i}k_{i}=0
\end{equation}
 Eqn.\ref{red_7} is the fundamental equation of geometrical optics and called the eikonal equation. Using Eqn.\ref{red_6} the components of the wave four-vector, $k_{i}$ 
are given as:
\begin{eqnarray}    
\begin{array}{lcl}
  k_{0}=-\frac{\partial \Psi}{\partial x^{0}}=-\frac{\partial \Psi}{\partial ct}=\frac{\nu^{\prime}}{c}\\
  k_{1}=-\frac{\partial \Psi}{\partial x^{r}}\\
  k_{2}=-\frac{\partial \Psi}{\partial x^{\theta}}\\
  k_{3}=-\frac{\partial \Psi}{\partial x^{\phi}}\\
\end{array}
\label{red_8}
\end{eqnarray}
$\nu$ is the frequency measured in terms of proper time, $\tau_{0} $ and is defined as\cite{landau1971classical}[page no. 268] $\nu=-\frac{\partial \Psi}{\partial \tau_{0}}$. 
Now, one must note that frequency $\nu $ expressed in terms of proper time $\tau_{0}$, is different at different point of space. $\nu^{\prime}$ is the frequency as  measured by the 
rotating observer. In the present case of interest where the light ray is coming from distance inertial frame and   observed by  rotating ( non-inertial) observer situated 
at the surface of radius of rotating frame, we can write from the Eqn.\ref{red_7} (please also see \cite{dubey2015analysis}):
\begin{equation}\label{red_9}
  [k_{i}u^{i}]_{source}=[k_{0}u_{0}\left(
                                        1+\frac{k_{\phi}}{k_{0}}\frac{u^{\phi}}{u^{0}}
                                    \right)]_{source}
\end{equation}
and
\begin{equation}\label{red_10}
  [k_{i}u^{i}]_{observer}=[k_{0}u_{0}\left(
                                        1+\frac{k_{\phi}}{k_{0}}\frac{u^{\phi}}{u^{0}}
                                    \right)]_{observer}
\end{equation}
Now for the distance source $u^{0}_{source}=1$ and $u^{\phi}_{source}=0$, therefore
\begin{equation}\label{red_11}
  \left[
    k_{0}u^{0}
  \right]_{source}=\left[
                            k_{0}u^{0}\left(
                                        1+\frac{k_{\phi}}{k_{0}}\frac{u^{\phi}}{u^{0}}
                                    \right)
                    \right]_{observer}
\end{equation}
Now, for the observer $\frac{d\phi}{dt}= \Omega$, ( rotational velocity of observer's frame) $[u^{0}]_{observer}$ is given by,(from Eqn.\ref{red_3})
\begin{equation} \label{red_12}
  [u^{0}]_{observer}=\frac{1}{\sqrt{
  [1-\frac{\Omega^{2}r^{2}}{c^{2}}(1+\sin \theta \cos \phi)- \frac{\Omega^{2}r^{2}}{c^{2}}\sin \theta \sin \phi]+2\frac{\Omega^{2}r^{2}}{c^{2}}\sin^{2} \theta
  }}
\end{equation}
Noting that $\frac{u^{\phi}_{observer}}{u^{0}_{observer}}=\frac{\Omega}{c}$ from Eqn.\ref{red_11} we can have:

\begin{equation}\label{red_13}
  \nu^{\prime}\left[
                    \frac{1+\frac{k_{\phi}}{k_{0}}\frac{\Omega}{c}}{\sqrt{
  [1-\frac{\Omega^{2}r^{2}}{c^{2}}(1+\sin \theta \cos \phi)- \frac{\Omega^{2}r^{2}}{c^{2}}\sin \theta \sin \phi]+2\frac{\Omega^{2}r^{2}}{c^{2}}\sin^{2} \theta
  }}
                \right]=\nu
\end{equation}
At the location of the observer the observed frequency is defined by $\nu^{\prime}$. To calculate the redshift, one must calculate the value of $[\frac{k_{\phi}}{k_{0}}]_{observer}$. 
For a rotating frame like earth, the relativistic action, $S$ for a particle can be expressed as\cite{landau1971classical,dubey2015analysis}, $S=-Et+L\phi +S_{r}+ S_{\theta}$, where 
$E$ is the conserved energy and $L$ is the components of angular momentum around the axis of symmetry. Now the four momentum is $p_{i}=-\frac{\partial S}{\partial x^{i}}$ and 
for the propagation of photon we can replace action $S$ by the eikonal, $\Psi$ . Finally, we can write
\begin{eqnarray}    
\begin{array}{lcl}
   k_{0}=-\frac{\partial \Psi}{\partial x^{0}}=\frac{E}{c}\\
   k_{\phi}=-\frac{\partial \Psi}{\partial \phi}=-L
 \end{array}
\label{red_14}
\end{eqnarray}
Again, in case of photon the energy $E$ and linear momentum $p$ are expressed as, $E=pc$ considering the rest mass of photon is zero. The angular momentum about the rotation axis
of observer can be expressed as\cite{dubey2015analysis}, $L=pr\sin \theta \sin \phi$. So the value of $\frac{k_{\phi}}{k_{0}}$ is $r\sin \phi \sin \theta$. So from Eqn. \ref{red_13} 
it can be written:
\begin{equation}\label{red_15}
   \nu^{\prime}\left[
                    \frac{1+r\sin \theta \sin \phi \frac{\Omega}{c}}{\sqrt{
  [1-\frac{\Omega^{2}r^{2}}{c^{2}}(1+\sin \theta \cos \phi)- \frac{\Omega^{2}r^{2}}{c^{2}}\sin \theta \sin \phi]+2\frac{\Omega^{2}r^{2}}{c^{2}}\sin^{2} \theta
  }}
                \right]=\nu
\end{equation}
The redshift, $Z$ is generally defined by the relation. $Z=\frac{\nu^{\prime}}{\nu}-1$. So the redshift as estimated by a rotating observer will be  given by:
  \begin{equation}\label{red_16}
   Z=\frac{\nu^{\prime}}{\nu}-1=\frac{{\sqrt{
  [1-\frac{\Omega^{2}r^{2}}{c^{2}}(1+\sin \theta \cos \phi)- \frac{\Omega^{2}r^{2}}{c^{2}}\sin \theta \sin \phi]+2\frac{\Omega^{2}r^{2}}{c^{2}}\sin^{2} \theta
  }}}{1+r\sin \theta \sin \phi \frac{\Omega}{c}}-1
 \end{equation}
Now when $\frac{\nu^{\prime}}{\nu} < 1$ then we call it redshift and when $\frac{\nu^{\prime}}{\nu} > 1$ we call it blueshift. From Eqn.\ref{red_16} one can observe that for 
equatorial plane there will be no shift of wavelength when the azimuthal position $(\phi)$ is zero, but the shift will be extremum at $\phi= \frac{\pi}{2}$ (dusk side) and
$\phi= -\frac{\pi}{2}$ (dawn side). We know for earth  $\Omega_{earth}$ is $7.27 \times 10^{-5}$ rad/sec and the average radius of earth is $r_{earth}$ is $63.78 \times 10^{5}$ m. 
Therefore, the redshift values under different conditions will be : $Z_{earth, \phi=\frac{\pi}{2}, \theta=\frac{\pi}{2}}\simeq-1.604429\times 10^{-6}$, i.e. redshift occurs 
here at dusk. Again for the azimuthal position $\phi=-\frac{\pi}{2}$ (i.e. at dawn )on the equatorial plane of earth, $Z$ will be $Z_{earth, \phi=-\frac{\pi}{2},
\theta=\frac{\pi}{2}}\simeq1.604436\times 10^{-6}$ which shows a blueshift. Now if we vary the polar position of the observer then at both the poles there will be no shift of 
the wavelength of the received light. Again from Eqn. (\ref{red_16}) one can find that for both cases when either $\theta$ or $\phi$ becomes zero the value of $Z$ will become zero. 
From our above discussion it is clear that the redshift attains extremum value when $\theta=\frac{\pi}{2}$ and $\phi= -\frac{\pi}{2}$ or $\frac{\pi}{2}$. Moreover it can be said 
that redshift is maximum at $\phi=\frac{\pi}{2}, \theta=\frac{\pi}{2}$ and blushift is maximum at $\phi=-\frac{\pi}{2}, \theta=\frac{\pi}{2}$.\\
  Let us consider a light ray consisting of $Lyman-\alpha$ line which has frequency, 
  $\nu_{Ly-\alpha}= 2.47 \times 10^{15} Hz$ ($\lambda_{Ly-\alpha}=1215.673123{\buildrel _{\circ} \over {\mathrm{A}}} $) and is
emitted from some distance inertial frame. Now, at dawn ($\phi=-\frac{\pi}{2}$) the received frequency of 
$Ly-\alpha$ will be $\nu^{\prime}_{-\frac{\pi}{2}}=2.470004\times 10^{15}Hz$ ($\lambda^{\prime}_{-\frac{\pi}{2}}=1215.671178 {\buildrel _{\circ} \over {\mathrm{A}}}$), 
again at dusk ($\phi=\frac{\pi}{2}$) the received frequency of 
$Ly-\alpha$ will be $\nu^{\prime}_{\frac{\pi}{2}}=2.46999\times 10^{15}Hz$ ($\lambda^{\prime}_{\frac{\pi}{2}}=1215.675068 {\buildrel _{\circ} \over {\mathrm{A}}}$). 
But there will be no shift at mid-noon ($\phi =0$). Thus we see the difference in the third place after decimal when wavelength is expressed in Angstroms. A present 
day high resolution astronomical spectrometer can resolve spectral lines with milli-Angstroms resolutions.  So in principle one can verify the predictions of our analysis 
on the redshift calculations . 
\subsection{Relativistic Doppler Shift for rotating frame}\label{subsec:doppler}
  As in the case for rotating observer, it is clear that the observer is either moving towards the source or away from the source, so this causes a
  relativistic Doppler effect. In this present work let us consider the observer is on the rotating frame which receives a light ray with wavelength $\lambda^{\prime}_{Dop}$ 
  from a distance source which initially emits a light ray with wavelength $\lambda_{Dop}$. The two wavelengths are related by the relation in case of relativistic Doppler effect 
  \cite{born2013principles}:
 \begin{equation} \label{doppler_1}
      \frac{\lambda_{Dop}}{\lambda^{\prime}_{Dop}}=\gamma (1-\beta)
  \end{equation}
 where $\beta =\frac{v}{c}$, $v $ is the translational velocity of the observer parallel to $x^{1}$-axis. Here as the observer is sitting on the rotating frame of radius $r$ and 
 the polar and azimuthal positions are given by $\theta$ and $\phi$ respectively, the velocity will be $\gamma^{2} \Omega r\sin \theta \cos \phi$.
 We note that $\gamma$ is the Lorentz factor and $\Omega$ is the angular velocity of the rotating frame. So the $\beta $ is given by the relation
 $\beta=\gamma^{2} \frac{\Omega r}{c}\sin \theta \cos \phi$. The redshift related to the relativistic Doppler effect is:
  \begin{equation}\label{doppler_2}
  Z_{Dop}=\frac{\lambda^{\prime}_{Dop}}{\lambda_{Dop}}-1=\frac{1}{\gamma (1-\gamma^{2} \frac{\Omega r}{c}\sin \theta \cos \phi)}-1
  \end{equation}
  A simple mathematical analysis will show that, the results obtained in Eqn.\ref{red_15} and Eqn.\ref{doppler_1} are essentially the same. 

The Eqn.\ref{red_15} was obtained through a rigorous procedure involving metric for non-inertial accelerated (rotational)  frame. 
  These results for the relativistic redshift are same as the Doppler redshift derived considering the rotational frame. Thus the redshift effect for the rotational frame is 
  equivalent to the relativistic redshift caused by the translational motion of the observer.
   
\section{Conclusion} \label{section:conclutsion}
In this paper, we studied the change in the polarization vector of light, coming from a distant inertial source and as viewed by a non-inertial  observer sitting on a rotating frame. 
The amount of rotation primarily depends on the four variables, i.e. (r, $\theta$, $\phi$) coordinate of the rotating observer, and the rate of rotation $\Omega $, of the coordinate frame. 
The maximum change of polarization vector can be observed, if the position of the observer is on the equatorial plane and the azimuthal position is $\pm \frac{\pi}{2}$ on the rotating frame. 
As an extension of the present work, the earth has been taken as an example of a rotating frame where the observer is on the surface of the earth and the maximum change in 
the direction of polarization vector, occurs at latitude $0^{\circ}$ ($ \theta =90^{\circ}$), and longitude $\pm 90^{\circ}$. This maximum value  is 
$\chi_{earth,\theta=\phi=\frac{\pi}{2}}= 0.33$ arc-sec. Again if we consider that the rotating axis of the frame is along the direction of propagation of light, then there will be 
no change in the direction of the polarization vector.\\  

In the present paper, we also report the amount of the redshift  as observed an observer on a  rotating frame. It has been shown that there will be redshift in dusk (prograde) case and 
will be blueshift for the dawn (retrograde) case (i.e. maximum redshift at azimuthal position $\phi=\frac{\pi}{2}$ on the equatorial plane and maximum blueshift at azimuthal position 
$\phi=-\frac{\pi}{2}$). We calculate the red shifted and blue shifted values of wavelength taking Lyman -alpha line as a test case. It is also shown that results from our 
calculations considering metric tensors for an accelerated (rotational) frame, come out to be same as the one which can be obtained considering the expression for relativistic 
Doppler shift.  

From above discussion, it is clear that both the changes in the direction of polarization vector ($\chi$) and redshift ($Z$) are maximum at the azimuthal position,
$\phi=\frac{\pi}{2}$ and polar position, $\theta=\frac{\pi}{2}$. Both $\chi$ and $Z$ become zero when either one of the coordinates  $\phi$ and $\theta$ becomes zero or $180^{\circ} $. 
Again one must notice that the value of $\chi$ and $Z$ in the azimuthal positions $\frac{\pi}{2}$ and $-\frac{\pi}{2}$ have the same mod value but with opposite sign.  

\begin{acknowledgements}
The authors would like to thank Dr. Atri Deshmukhya, Head of the Department of Physics, Assam University, Silchar, India for useful discussions and suggestions. T.Ghosh is also thankful to Dr. Anuj Kumar Dubey, and Amritaksha Kar Department of Physics and Hirak Chatterjee Department of Chemistry, Assam University, for providing help and support in programming and making plots. 
\end{acknowledgements}


%

\bibliography{aipsamp}

\end{document}